\documentclass[preprint,prc,aps,epsfig]{revtex4}
\usepackage{psfig}
\def\beq{\begin{equation}}
\def\enq{\end{equation}}
\def\beqa{\begin{eqnarray}}
\def\enqa{\end{eqnarray}}

\def\GeV{\nobreak\,\mbox{GeV}}

\def\pli{p^\prime}
\def\mli{{M^\prime}^2}

\def\la{\lambda}
\def\ga{\gamma}
\def\Ga{\Gamma}

\def\si{\sigma}

\def\al{\alpha}

\def\lb{\label}
\def\nn{\nonumber}
\newcommand{\rag}{\rangle}
\newcommand{\lag}{\langle}

\newcommand{\rf}{\ref}
\newcommand{\ct}{\cite}

\begin{document}

\title{\sc  $D_s$ decays into $\phi$ and $f_0(980)$ mesons}
\author { Ignacio Bediaga$^1$ and Marina Nielsen$^2$}
\affiliation{ $^1$Centro Brasileiro de Pesquisas F\'\i sicas\\
Rua Xavier Sigaud 150, 22290-180 Rio de Janeiro, RJ, Brazil\\
$^2$Instituto de F\'{\i}sica, Universidade de S\~{a}o Paulo\\
  C.P. 66318,  05315-970 S\~{a}o Paulo, SP, Brazil}

\begin{abstract}
We consider the nonleptonic and semileptonic decays of $D_s$-mesons into 
$\phi$ and $f_0(980)$ mesons. QCD sum rules are used to calculate the 
form factors associated with these decays, and the correspondig decay 
rates.
On the basis of data on $D_s^+\rightarrow\pi^+\pi^+\pi^-$, which goes
dominantly via the transition $D_s^+\rightarrow \pi^+f_0(980)$, we conclude 
that there is space for a sizeable light quark component on $f_0(980)$.
\end{abstract}

\maketitle

\vspace{1cm}
\section{Introduction}

The interpretation of the nature of the
lightest scalar mesons have been controversial
since their first observation over thirty years ago. Due to the complications
of the nonperturbative strong interactions there is still no general
agreement about their structure. Actually, the observed light scalar states
are too numerous \cite{PDG} to be accomodated in a single $q\bar{q}$ 
multiplet, and therefore, it has been suggested that some of 
them escape the quark model interpretation. It is not known 
whether there is necessarily
a glueball among the light scalar, and whether some of the too numerous scalars
are multiquark or some meson-meson bound states, or even admixtures of
quarks and gluons \cite{cloto}.

In particular, the structure of the meson $f_0(980)$ has been extensively
debated. It has been interpreted as an $s\bar{s}$ state \cite{torn,bev,col},
as an four quark $s\bar{s}q\bar{q}$ state \cite{jaffe}, as a bound state
of hadrons \cite{wein,clo1}, and as a result of a process known as hadronic 
dressing {\cite{torn,bev,faz}.

The recently measured relative weight of the reaction 
$D_s^+\rightarrow f_0(980)\pi^+\rightarrow\pi^+\pi^-\pi^+$ \cite{Ait01}, 
may serve as a tool for the estimation of the $s\bar{s}$
component of the meson $f_0(980)$. As a matter of fact, if $f_0(980)$ has a 
pure strangeness component ($f_0(980)=s\bar{s}$), the dominant 
$D_s^+\rightarrow 
f_0(980)\pi^+$ decay proceeds via the spectator mechanism, as shown in Fig.~1.
However, in the four quark scenario ($f_0(980)=s\bar{s}(u\bar{u}+d\bar{d})/2$)
the decay $D_s^+\rightarrow f_0(980)\pi^+$ is expected to proceed through a 
much more complicated recombination. 

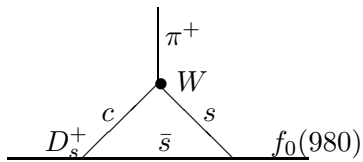
\begin{figure}[h] 
\setlength{\unitlength}{0.5cm}
\begin{center} \begin{picture}(8,5) \put(2,0){\line(1,1){2}}
\put(4,2){\line(1,-1){2}} \put(2,0){\line(1,0){4}} 
\put(0,0){\line(1,0){2}}
\put(4,2){\line(0,1){2}}
\put(6,0){\line(1,0){2}}
\put(1,0.2){$D_s^+$}
\put(7,0.2){$f_0(980)$}
\put(4.2,3){$\pi^+$}
\put(4,0.2){$\bar{s}$}
\put(4.5,1.8){$W$}
\put(3.9,1.8){$\bullet$}
\put(2.5,1){$c$}
\put(5.2,1){$s$}
\end{picture}
\end{center}
\vspace{0.5cm}
\caption{Schematic picture  of the spectator mechanism for the decay
$D_s^+\rightarrow f_0(980)\pi^+$.} \lb{graph} 
\end{figure}

Since the spectator mechanism provides a strong production of the
$\phi(1020)$ meson in the decay $D_s^+\rightarrow \phi\pi^+$, in this work
we consider the ratio
\beq
R={\Ga(D_s^+\rightarrow f_0(980)\pi^+)\over \Ga(D_s^+\rightarrow 
\phi\pi^+)}\;,
\label{rat}
\enq
to evaluate the importance of the $s\bar{s}$ component in the $f_0(980)$ meson.
This same ratio was evaluated in recent calculations  by using the
spectral integration technique \cite{adn}, and the constituent quark meson
model \cite{detc}. In both calculations the authors concluded that the
$s\bar{s}$ component dominates the $f_0(980)$ meson and, therefore,
the spectator mechanism dominates the $D_s^+\rightarrow f_0(980)\pi^+$ decay.
Here we use the QCD sum rules method to evaluate the ratio in Eq.~(\ref{rat}),
as well as the branching ratios for the nonleptonic $D_s^+\rightarrow 
\phi\pi^+$ and semileptonic $D_s^+\rightarrow \phi\ell^+ \nu_\ell$ and
$D_s^+\rightarrow f_0(980)\ell^+ \nu_\ell$ decays.
The two branching ratios involving the meson $\phi$ will be used to check 
the reliability of
the method, since these two branching ratios are known experimentaly
\cite{PDG}:
\beq
{\cal B}^{exp}(D_s^+\rightarrow \phi\pi^+)=(3.6\pm0.9)\%\;,
\label{phiexp}
\enq
\beq
{\cal B}^{exp}(D_s^+\rightarrow \phi\ell^+ \nu_\ell)=(2.0\pm0.5)\%\;.
\label{philep}
\enq

\section{Decay widths}

The decay width of the nonleptonic process $D_s^+\rightarrow M\pi^+$, where $M$
stands for the $\phi$ or $f_0(980)$ mesons, is given by:
\beq
\Ga(D_s^+\rightarrow M\pi^+)={1\over16\pi m_{D_s}^3}|{\cal M}|^2
\sqrt{\la(m_{D_s}^
2,m_M^2,m_\pi^2)}\;,
\label{wid}
\enq
with $\la(x,y,z)=x^2+y^2+z^2-2xy-2xz-2yz$. The QCD factorization formula
(in the limit $m_\pi^2\rightarrow0$) gives for $f_0(980)$:
\beq
|{\cal M}(D_s^+\rightarrow f_0(980)\pi^+)|^2={G_F^2\over2} |V_{cs}|^2|V_{ud}|^2
\left(c_1+{c_2\over 3}\right)^2f_\pi^2 (m_{D_s}^2-m_{f_0}^2)^2f_+^2(0),
\label{amf0}
\enq
where $f_+$ is the $D_s\rightarrow f_0(980)$ form factor defined as
\beq
\lag f_0(\pli)|\bar{s}\ga_\mu(1-\ga_5)c|D_s(p)\rag=i(f_+(t)(p+\pli)_\mu
+f_-(t)q_\mu),
\label{faf0}
\enq
with $t=q^2$ and $q=p-\pli$. And for $\phi$ we have
\beqa
|{\cal M}(D_s^+\rightarrow \phi\pi^+)|^2&=&{G_F^2\over8m_\phi^2} |V_{cs}|^2
|V_{ud}|^2\left(c_1+{c_2\over 3}\right)^2f_\pi^2 \la(m_{D_s}^2,m_\phi^2,
m_\pi^2)[(m_{D_s}+m_\phi)A_1(0)
\nonumber\\
&-&(m_{D_s}-m_\phi)A_2(0)]^2,
\label{amphi}
\enqa
where $A_1$ and $A_2$ are  $D_s\rightarrow \phi$ form factors defined as
\cite{bbd91}
\beqa
\lag \phi(\pli)|\bar{s}\ga_\mu(1-\ga_5)c|D_s(p)\rag&
=&-i(m_{D_s}+m_\phi)A_1(t)\epsilon^\phi_\mu+
i{A_2(t)\over m_{D_s}+m_\phi}\epsilon^\phi.p(p+\pli)_\mu
\nonumber\\
&+&i{A_3(t)\over m_{D_s}+
m_\phi}\epsilon^\phi.p~ q_\mu+{2V(t)\over m_{D_s}
+m_\phi}\epsilon_\mu^{\nu\rho
\sigma}\epsilon^\phi_\nu p_\rho\pli_\sigma.
\label{faphi}
\enqa

In Eqs.~(\ref{amf0}) and (\ref{amphi}) the coefficients $c_1$ and
$c_2$ are the Wilson coefficients entering the effective weak Hamiltonian
evaluated at the normalization scale $\mu$ \cite{bubu}:
\beq
H_W={G_F\over\sqrt{2}}V_{cs}V^*_{ud}\left[\left(c_1(\mu)+{c_2(\mu)\over 3}
\right)O_1+...\right],
\enq
with $O_1=(\bar{u}\ga_\mu(1-\ga_5)d)(\bar{s}\ga^\mu(1-\ga_5)c)$.
Therefore, in calculating the ratio in Eq.~(\ref{rat}) we are free from the
uncertainties in the Wilson coefficients and in the CKM transition
elements.

In the case of the semileptonic decays $D_s^+\rightarrow M\ell^+\nu_\ell$
the differential decay rates are given by
\beq
\frac{d\Ga}{dt} =
\frac{G_F^2|V_{cs}|^2}{192\pi^3m_{D_s}^3} \la^{3/2}(m_{D_s}^2,m_{f_0}^2,t)
f_{+}^2(t) \;,
\lb{lepf0}
\enq
for $D_s^+\rightarrow f_0(980)\ell^+\nu_\ell$. The decay rate for the decay
$D_s^+\rightarrow \phi\ell^+\nu_\ell$ is written in terms of the
helicity amplitudes
\beq
H_\pm(t)=(m_{D_s}+m_\phi)A_1(t)\mp{\la^{1/2}(m_{D_s}^2,m_\phi^2,t)\over
m_{D_s}+m_\phi}V(t)\;,
\enq
\beq
H_0(t)={1\over2m_\phi\sqrt{t}}\left((m_{D_s}^2-m_\phi^2-t)
(m_{D_s}+m_\phi)A_1(t)-{\la(m_{D_s}^2,m_\phi^2,t)\over
m_{D_s}+m_\phi}A_2(t)\right)\;,
\enq
so that
\beq
\frac{d\Ga_\pm}{dt} =
\frac{G_F^2|V_{cs}|^2}{192\pi^3m_{D_s}^3}t \la^{1/2}(m_{D_s}^2,m_{f_0}^2,t)
|H_\pm(t)|^2\;,
\enq
\beq
\frac{d\Ga_L}{dt} =
\frac{G_F^2|V_{cs}|^2}{192\pi^3m_{D_s}^3}t \la^{1/2}(m_{D_s}^2,m_{f_0}^2,t)
|H_0(t)|^2\;,
\enq
\beq
\frac{d\Ga_T}{dt} =\frac{d}{dt}(\Ga_++\Ga_-),\;\;\;\; \frac{d\Ga}{dt} =
\frac{d}{dt}(\Ga_L+\Ga_T)\;.
\lb{lepphi}
\enq

\section{Sum rules}

The $D_s^+$ meson in the initial state is interpolated by the pseudoscalar
current
\beq
j_{D_s}(x) = \bar{s}(x)i\ga_5 c(x)\;,
\lb{intD}
\enq
where $c$ and $s$ are the fields of the charmed and strange quark 
respectively. Summation over spinor and colour indices
being understood but not indicated explicitly. The final hadronic  state $M$
is interpolated by the current 
\beq
j_M(x) = \bar{s}(x)\Ga_M s(x)\;,
\lb{intscal}
\enq
where 
\beq
\begin{array}{c}
\Ga_M=\end{array}
\left\{
\begin{array}{c}
1\mbox{ for }M=f_0(980)\\
\ga_\al \mbox{ for }M=\phi \end{array}
\right.
\enq

Using the QCD sum rule technique \cite{svz}, the form factors in 
Eqs.~(\ref{faf0}) and (\ref{faphi}) can be evaluated from the time ordered 
product of the two interpolating fields in Eqs.~(\rf{intD}) and 
(\rf{intscal}) and the weak current $j^W_\mu=\bar{s}\gamma_\mu(1-\gamma_5) c$
\beq
T_{\mu M}(p,\pli) = i\int d^4x d^4y\, \lag0|{\rm T}[ j_M(x) j^W_\mu(y)
j_{D_s}^\dagger(0)]|0\rag e^{i(\pli.x-q.y)}\;.
\lb{3point}
\enq

In order to evaluate the phenomenological side
we insert intermediate states for $D_s$ and $M$, we use the definitions 
in Eqs.~(\ref{faf0}) and (\ref{faphi}), 
and obtain the following relations
\beqa
T_{\mu}^{phen} (p,\pli)&=&m_{f_0}f_{f_0}{m_{D_s}^2f_{D_s}\over m_c+m_s}
~{f_+(t)(p+\pli)_\mu+f_-(t)~q_\mu\over(p^2-m_{D_s}^2)({\pli}^2-m_{f_0}^2)}
\nn\\
&+&
\mbox{ contributions of higher resonances}\;,
\lb{pf0}
\enqa
for $f_0(980)$, and
\beqa
T_{\mu\al}^{phen} (p,\pli)&=&m_\phi f_\phi{m_{D_s}^2f_{D_s}\over m_c+m_s}
{1\over(p^2-m_{D_s}^2)({\pli}^2-m_\phi^2)}\bigg(-(m_{D_s}+m_\phi)A_1(t)
g_{\mu\al}
\nn\\
&+&{A_2(t)\over (m_{D_s}+m_\phi)}(p+\pli)_\mu~p_\al
-2i{V(t)\over (m_{D_s}+m_\phi)}
\epsilon_{\mu\al\rho\si}~p_\rho\pli_\si
\nn\\
&+&\cdots \bigg)+
\mbox{ contributions of higher resonances}\;,
\lb{pphi}
\enqa
for $\phi$, where we have shown only the structures important for the 
evaluation of the form factors $A_1,~A_2$ and $V$.

In the above equations the coupling of the $f_0(980)$ to the scalar
current $j_s=\bar{s}s$, was parametrized in terms of the constant
$f_{f_0}$ as:
\beq
\lag 0 |\bar{s}s|f_0\rag =m_{f_0}~f_{f_0}\;,
\lb{f0}
\enq
and we have used the standard definitions of the couplings of $D_s$ and $\phi$
with the corresponding currents:
\beq
\lag 0 | j_{D_s}|D_s\rag ={m_{D_s}^2f_{D_s}\over m_c+m_s}\;,
\lb{ds}
\enq
\beq
\lag 0 |\bar{s}\ga_\al s|\phi\rag =m_\phi~f_\phi\epsilon_\al\;,
\lb{phi}
\enq

The three-point function Eq.(\ref{3point}) can be
evaluated by perturbative QCD if the external momenta are in the deep Euclidean
region
\beq
p\ll (m_c+m_s)^2,~~~ {\pli}^2 \ll 4m_s^2, ~~~ t \ll (m_c+m_s)^2\;.
\lb{cond} 
\enq
In order to approach the not-so-deep-Euclidean region and to 
get more information on the nearest physical singularities, 
nonperturbative power corrections are added to the perturbative contribution.
In practice, only the first few condensates contribute significantly, the
most important ones being the 3-dimension, $\lag\bar{s}s\rag$,
and the 5-dimension, $\lag\bar{s}g_s\si.Gs\rag$, condensates.
For each invariant structure, $i$, we can write
\beqa
T^{theor}_{i}(p^2,{\pli}^2,t)&=& {-1\over4\pi^2}\int_{(m_c+m_s)^2}^\infty ds
\int_{4m_s^2}^\infty du\,
\frac{\rho_i(s,u,t)}{(s-p^2)(u-{\pli}^2)}
\nn\\
&+& T^{D=3}_{i}\lag\bar{s}s\rag+ T^{D=5}_{i}\lag\bar{s}g_s\si.Gs\rag+\cdots\;.
\lb{power}
\enqa
The perturbative contribution is contained in the double discontinuity
$\rho_{i}$.

In order to suppress the condensates of higher dimension and at the same time
reduce the influence of higher resonances, the series in Eq.~(\rf{power}) is
Borel improved,  leading to the  mapping 
\beq
f(p^2) \to \hat f(M^2),~~~~~\frac{1}{(p^2-m^2)^n} \to \frac{(-1)^n}{(n-1)!}
\frac{e^{-m^2/M^2}}{(M^2)^n}\;.
\enq
Furthermore, we make the usual assumption that the contributions of  higher
resonances are well approximated by the perturbative expression
\beq
{-1\over4\pi^2}\int_{s_0}^\infty ds
\int_{u_0}^\infty du\,
\frac{\rho_i(s,u,t)}{(s-p^2)(u-{\pli}^2)}
\enq 
with appropriate continuum thresholds $s_{0}$ and $u_{0}$.
By equating the Borel transforms of the phenomenological expression
for each invariant structure in Eqs.(\rf{pf0}), (\rf{pphi}) and that of the 
``theoretical expression'', 
Eq.~(\rf{power}), we obtain the sum rules for the form factors (at the order 
$m_s$):
\beqa
&&
C_{f_0}e^{-m_{D_s}^2/M^2}
e^{-m_{f_0}^2/\mli}f_+(t)={-1\over4\pi^2}\int_{(m_c+ms)^2}^{s_0} ds 
\int_0^{u_0}du\left[e^{-s/M^2}e^{-u/\mli}\rho_+(s,u,t)\right]
\nonumber\\
&+&{\lag\bar{s}s\rag\over2}e^{-m_c^2/M^2}\left[-m_c+2m_s+{m_c^2m_s\over2M^2}
\right]+\lag\bar{s}g_s\si.Gs\rag e^{-m_c^2/M^2}\left[{m_c^2(m_c-m_s)\over8M^4}
-{2m_c-m_s\over6M^2}\right.\nonumber\\
&+&\left.{m_c^2(4m_c-3m_s)-2t(m_c-m_s)\over24M^2\mli}-{m_c-2m_s\over6\mli}
+{m_c^2m_s-2t(m_c-m_s)\over24M^2\mli}
\right],
\lb{f+}
\enqa
\beqa
&&C_\phi(m_{D_s}+m_\phi)^2e^{-m_{D_s}^2/M^2}e^{-m_\phi^2/\mli}A_1(t)=
{-1\over4\pi^2}\int_{(m_c+ms)^2}^{s_0} ds 
\int_0^{u_0}du\left[e^{-s/M^2}e^{-u/\mli}\right.\nonumber\\
&\times&\left.\rho_1(s,u,t)\right]
+{\lag\bar{s}s\rag\over2}e^{-m_c^2/M^2}\left[t-m_c^2-{5\over2}m_cm_s
-{m_cm_s\over2M^2}(t-m_c^2)\right]\nonumber\\
&+&\lag\bar{s}g_s\si.Gs\rag e^{-m_c^2/M^2}
\left[-{1\over6}
+{m_c^2(m_c^2+2m_cm_s-t)\over8M^4}
-{2m_c^2+3m_cm_s-2t\over12\mli}\right]
\nonumber\\
&-&\left.{3m_c^2+9m_cm_s-4t\over12M^2}
+{m_c(2m_c^3+3m_cm_s+tm_s)/2-t(2m_c^2+3m_cm_s/2-2t)\over6M^2\mli}
\right],
\lb{a1}
\enqa
\beqa
C_\phi e^{-m_{D_s}^2/M^2}e^{-m_\phi^2/\mli}A_2(t)&=&
{-1\over4\pi^2}\int_{(m_c+ms)^2}^{s_0} ds 
\int_0^{u_0}du\left[e^{-s/M^2}e^{-u/\mli}\rho_2(s,u,t)\right]\nonumber\\
&+&{\lag\bar{s}s\rag\over2}e^{-m_c^2/M^2}\left(1-{m_cm_s\over2M^2}\right)
-\lag\bar{s}g_s\si.Gs\rag e^{-m_c^2/M^2}
\nonumber\\
&\times&\left[{1\over6M^2}+
{m_c^2\over8M^4}+{2m_c^2-m_cm_s-2t\over12M^2\mli}\right],
\lb{a2}
\enqa
and 
\beqa
-2C_\phi e^{-m_{D_s}^2/M^2}e^{-m_\phi^2/\mli}V(t)&=&
{-1\over4\pi^2}\int_{(m_c+ms)^2}^{s_0} ds 
\int_0^{u_0}du\left[e^{-s/M^2}e^{-u/\mli}\rho_V(s,u,t)\right]\nonumber\\
&+&{\lag\bar{s}s\rag}e^{-m_c^2/M^2}\left(1-{m_cm_s\over2M^2}\right)
-\lag\bar{s}g_s\si.Gs\rag e^{-m_c^2/M^2}
\nonumber\\
&\times&\left[-{1\over3M^2}+
{m_c^2\over4M^4}+{2m_c^2-m_cm_s-2t\over6M^2\mli}\right],
\lb{v}
\enqa
where
\beq
C_{f_0}={m_{f_0}f_{f_0}m_{D_s}^2f_{D_s}\over m_c+m_s},\;\;
\mbox{and}\;\;C_\phi={m_\phi f_\phi m_{D_s}^2 f_{D_s}\over(m_c+m_s)
(m_{D_s}+m_\phi)}.
\enq
The decay constants $f_{D_s},\;f_{f_0}$ and $f_\phi$
defined in Eqs.~(\rf{f0}), (\rf{ds}), and(\rf{phi}), and appearing in the
constants $C_{f_0}$ and $C_\phi$, can also be determined  by sum rules
obtained from the appropriate two-point functions. The
explicit expressions for the two-point sum rules and for the
double discontinuities in Eqs.~(\rf{f+}),
(\rf{a1}), (\rf{a2}) and (\rf{v}) are given in  Appendices A and B 
respectively.

\section{Evaluation of the sum rules and results}

In the complete theory,  the form factors on the
right hand side of Eqs.~~(\rf{f+}),
(\rf{a1}), (\rf{a2}) and (\rf{v}) should not depend on the Borel
variables $M^2$ and $\mli$. However, in  a truncated treatment there will 
always  be some dependence left.  Therefore, one has to work in a region 
where the approximations made are supposedly acceptable and where 
the result depends only moderately on the Borel variables. 
To decrease the 
dependence of the results on the Borel variables $M^2$ , we take
them in the two-point functions at half the
value of the corresponding variables in the three-point sum rules 
\cite{bbd91,ra}. We furthermore choose
\beq
\frac{M^2}{\mli} = \frac{m_{D_s}^2-m_c^2}{m_M^2}\;. \lb{mm}
\enq
If the momentum transfer, $t$, is larger than a
critical  value $t_{cr}$,
non-Landau singularities have to be taken into account \ct{bbd91}. 
Since anyhow we have
to stay away from the physical region , {\it i.e.} we must have
$t \ll (m_c+m_s)^2$, we limit our calculation to the region $-0.5< t 
< 0.4\GeV$.
In this  range the t-dependence can be obtained from the sum rules 
directly. It can be fitted by a monopole, 
and extrapolated to the full kinematical region.

Since we do not take into account radiative corrections we choose the QCD
parameters at a fixed renormalisation scale of about $1~\GeV^2$:
the strange and
charm mass $m_s=0.14\,\GeV$, $m_c=1.3\,\GeV$.
We take for the strange quark condensate
$\langle\overline{s}s\rangle\,=0.8\langle\overline{q}q\rangle$ with
$\langle\overline{q}q\rangle\,=\,-(0.24)^3\,\GeV^3$, 
and for the  mixed quark-gluon condensate $\langle\overline{s}g_s\sigma.G
s\rangle=m_0^2 \langle\overline{s}s\rangle$ with 
$m_0^2=0.8\,\GeV^2$.

For the continuum thresholds we take the values discussed in the Appendix A:
$s_{0}=7.7\pm1.1~\GeV^2$ and,
\beq
\begin{array}{c}
u_0=\end{array}
\left\{
\begin{array}{c}
1.6\pm0.1\GeV\mbox{ in Eq.(\ref{f+}) }\\
2.0\pm0.1\GeV \mbox{ in Eqs.(\ref{a1}), (\ref{a2}) and (\ref{v}) }\end{array}
\right.
\enq

We evaluate our sum rules in 
the range $4.5\leq M^2\leq9.0\GeV^2$, which is compatible with the
Borel ranges used for the two-point functions in Appendix A.
In Fig.~2 we show the different  contributions to the form
factors $f_+,~A_1,~A_2$ and $V$ at zero momentum transfer, from the sum rules
in Eqs.~(\ref{f+}), (\ref{a1}), (\ref{a2}) and (\ref{v}),
as a function of the Borel variable
$M^2$, using  the continuum thresholds $s_{0}= 8.8 \GeV^2$ and 
$u_0=1.6\GeV^2$ or $2.0\GeV^2$ for $f_0$ or $\phi$ respectively. We see that 
$A_2(0)$ gets a big contribution from the quark condensate, while the 
perturbative
contribution is the largest one for all other form factors. Such kind
of behaviour had been already obtained in the $D\rightarrow K^*$
semileptonic decay studied in \cite{bbd91}. The mixed condensate contribution
is negligible for all four form factors, and the stability is quite
satisfactory in the Borel range studied. Varying the continuum
threshold $s_0$ in the range discussed in Appendix A, and also evaluating
the sum rules using or the expressions given in Appedix A for the meson decay 
constants, or its numerical values, we get for the form factors at $t=0$:
\begin{figure} \label{fig2}
\centerline{\psfig{figure=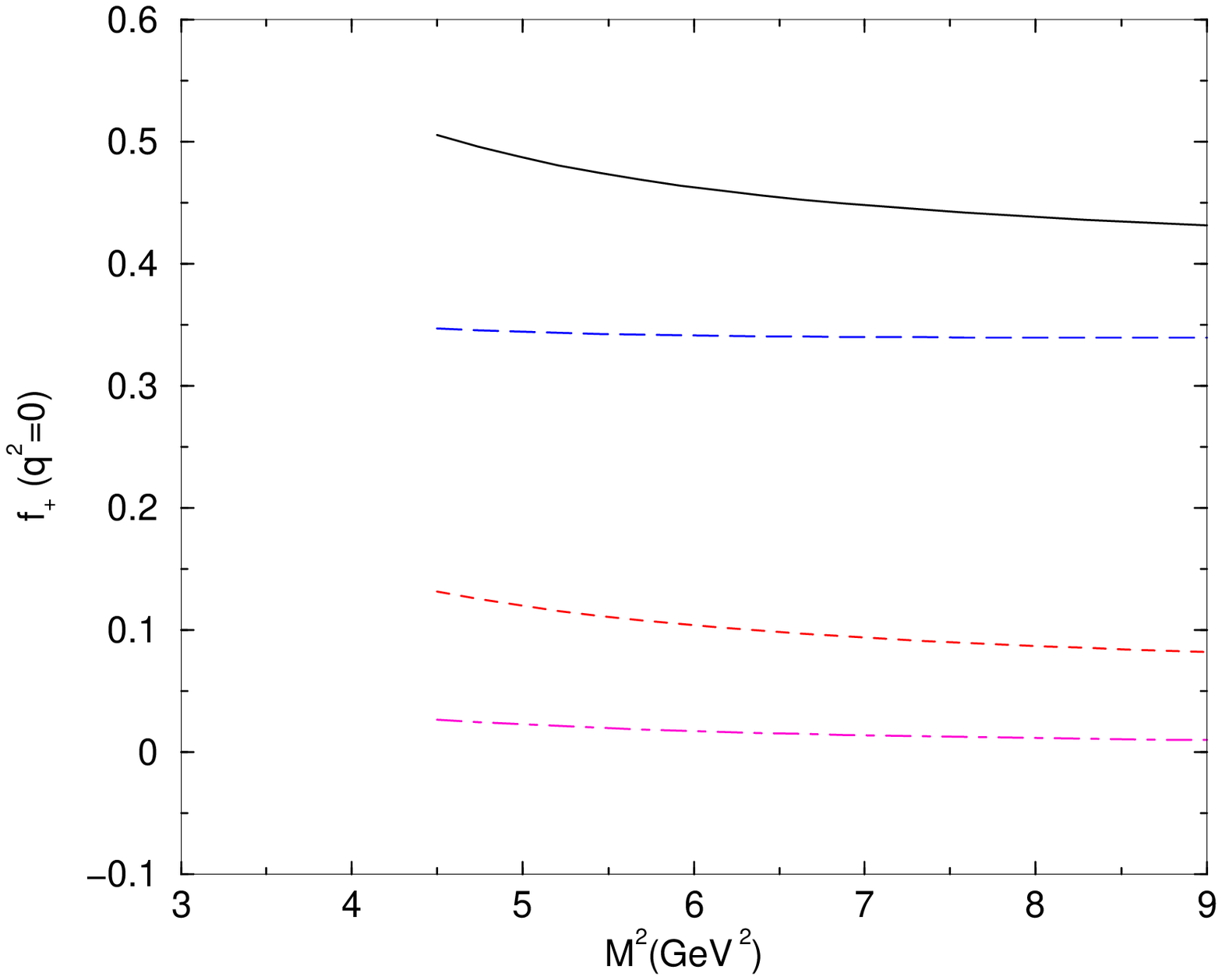,height=60mm,width=70mm,angle=0}
\psfig{figure=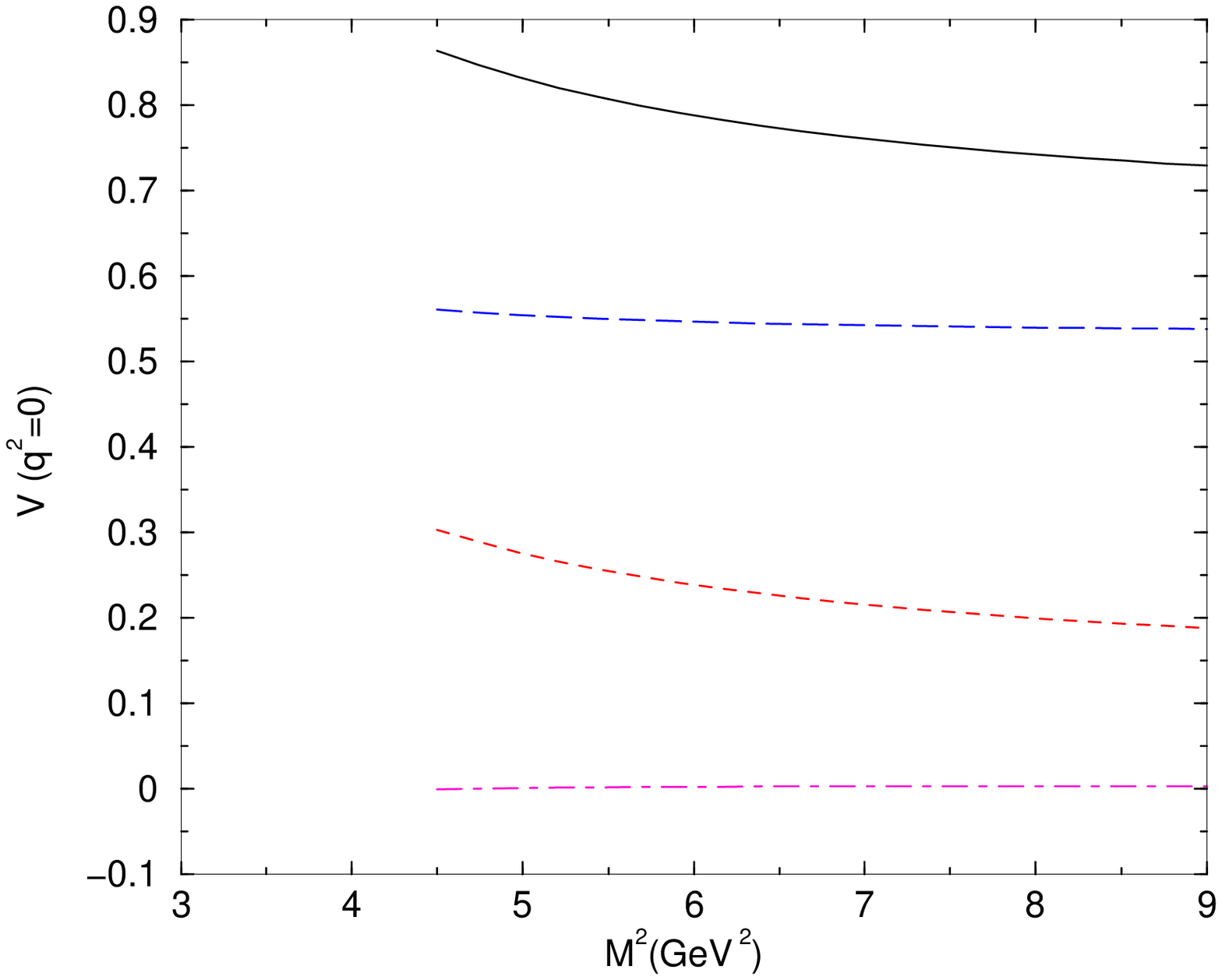,height=60mm,width=70mm,angle=0}}
\centerline{\psfig{figure=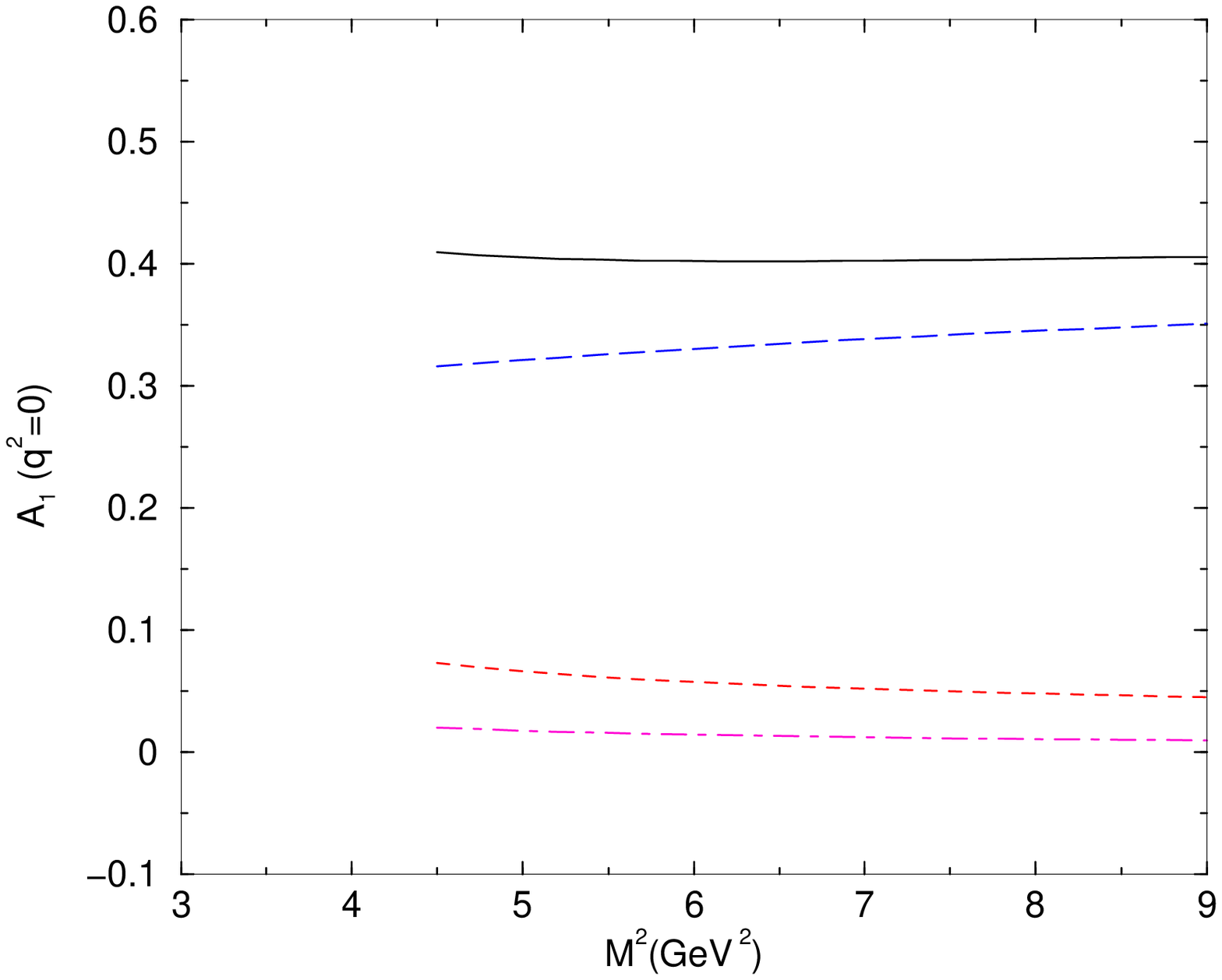,height=60mm,width=70mm,angle=0}
\psfig{figure=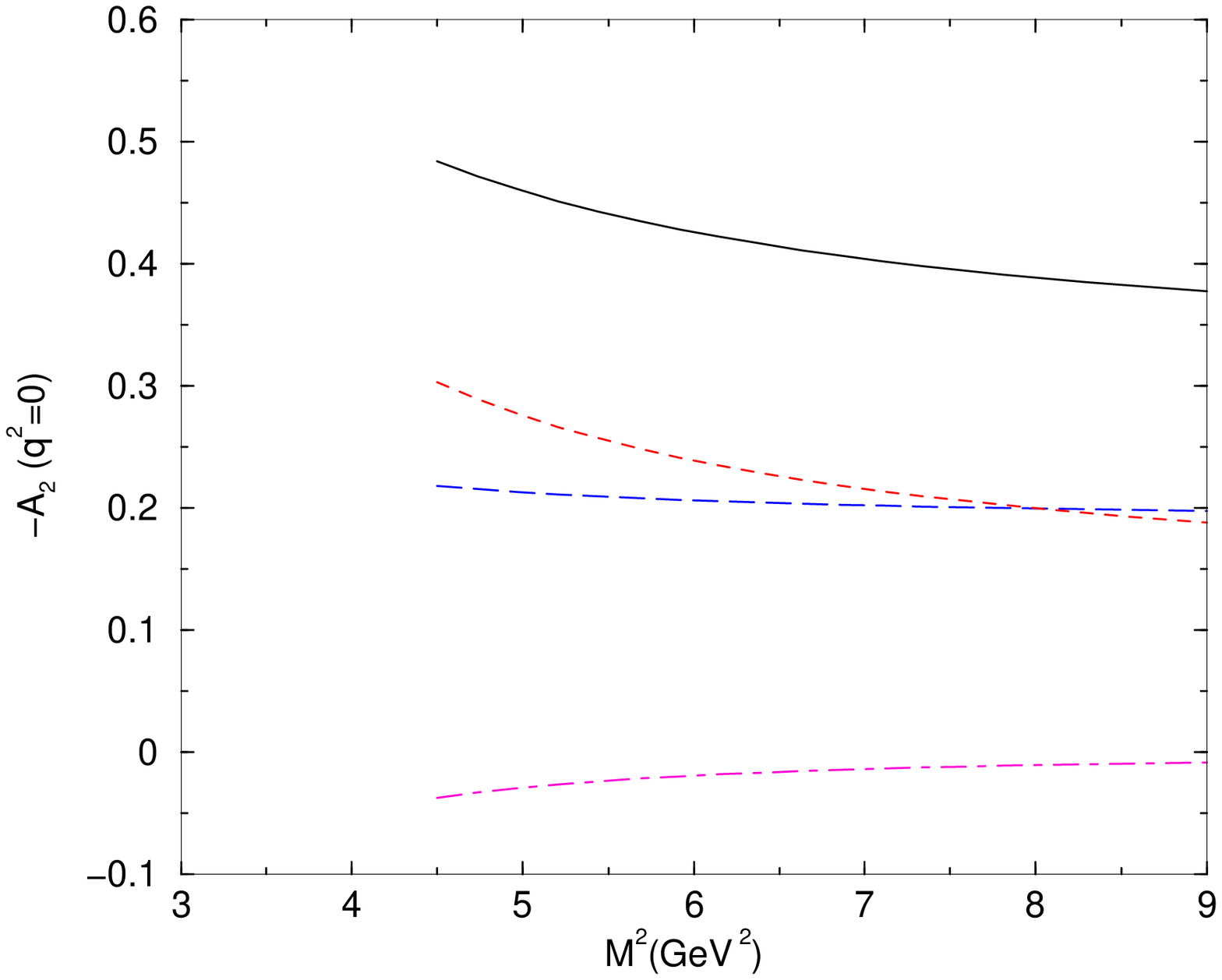,height=60mm,width=70mm,angle=0}}
\caption{Various contributions to the OPE of the form factors as a
function of the Borel parameter $M^2$.
 Solid curve: total contribution; long-dashed: perturbative; 
dashed: quark condensate; dot-dashed mixed condensate contribution.} 
\end{figure}
\beqa
0.40\leq f_+(0)\leq0.48,\;\;\;\;\;\;\;\;&&0.71\leq V(0)\leq0.89\;,\nn\\
0.32\leq A_1(0)\leq0.42,\;\;\;\;\;\;\;&-&0.43\leq A_2(0)\leq-0.37.
\lb{fornum}
\enqa
The value obtained for $f_+(0)$ is smaller than the value obtained for the 
same form factor in ref.~\cite{detc} by using the constituent quark meson 
model.

The $t$ dependence of the 
form factors evaluated at $M^2=7\GeV^2$ is shown in Fig.~3.
In the range  $-0.5\leq t \leq 0.4 \; \GeV^2 $ no non-Landau singularities
occur for our choices of the continuum thresholds.  The QCD sum rules 
results can, in this $t$-range, be very well  approximated by a monopole 
expression 
\beq
F(t)={F(0)\over 1-{t\over M_P^2}}\;,
\lb{mo}
\enq
for all four form factors. The reult of the fit is also shown
in Fig.~3. the different values for the pole mass, $M_P$, for the different
form factors are given by:
\beq
\begin{array}{c}
M_P=\end{array}
\left\{
\begin{array}{c}
(1.6\pm0.2)\GeV\mbox{ for }f_+(t) \\
(4.2\pm0.5)\GeV\mbox{ for }A_1(t) \\
(8.0\pm2.0)\GeV\mbox{ for }A_2(t) \\
(1.95\pm0.10)\GeV\mbox{ for }V(t) 
\end{array}
\right.
\enq

In the case of $A_2$ we get a very hight $M_P$ showing a very weak $t$
dependence. This approximate $t$ independence stems from a mutual cancelation
in the sum rule of an increase in the perturbative and a decrease in the
quark condensate contributions. Even for $A_1$ the $t$ dependence is much 
weaker than for $f_+$ and $V$. It is also interesting to notice
that $M_P^{(V)}$ is of the same order than the one for
the semileptonic $D\rightarrow K^*\ell \nu_\ell$ found in \cite{bbd91}, and 
$M_P^{(f_+)}$ is compatible with the ones found for the $D_s$ decays 
into $\eta$ \cite{cofa} and $D\to\kappa\ell \nu_\ell$ \cite{dfnn}.

\begin{figure} \label{fig3}
\centerline{\psfig{figure=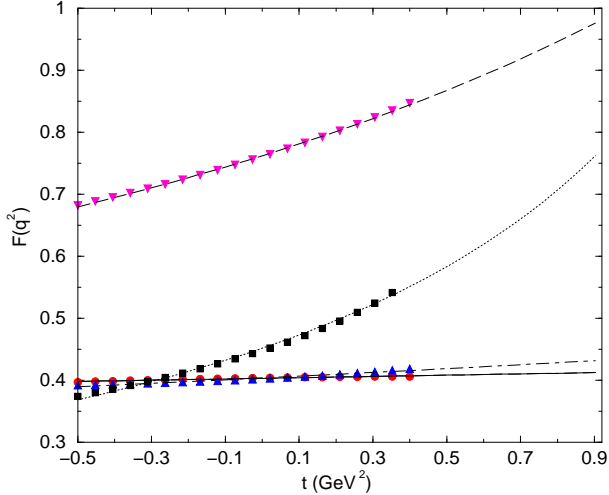,width=8cm,angle=0}}
\caption{$t$ dependence of the form factors. The solid, dot-dashed, dotted and
dashed lines give the monopole parametrization, in Eq.(\ref{mo}), to
the QCD sum rule results for $-A_2$ (circles), $A_1$ (triangles up),
$f_+$ (squares) and $V$ (triangles dow) respectively.}
\end{figure}

Having the form factors we can evaluate the decay widths for the
$D_s^+\rightarrow\phi\pi^+$ and $D_s^+\rightarrow \phi\ell^+ \nu_\ell$
decays, given by Eqs.~(\ref{wid}) and (\ref{lepphi}) respectively.
We obtain the following branching ratios:
\beq
{\cal B}(D_s^+\to\phi\pi^+)= (2.8 \pm 0.7)\% \;,
\enq
and
\beq
{\cal B}(D_s^+\to\phi\ell^+\nu)= (1.4 \pm 0.4)\% \;
\enq
where we have used $V_{ud}=V_{cs}=0.975$, $f_\pi=0.132$ and the values
$c_1(m_c)=1.263$ and $c_2(m_c)=-0.513$, corresponding to the results for
the Wilson coefficients obtained at the leading order in renormalization
group improved perturbation theory at $\mu\simeq1.3\GeV$ \cite{cofa}.
The errors in the above results were estimated only by taking into account
the uncertainties in the form factors in Eq.~(\ref{fornum}) and should
be understood as limiting values for the branching ratios.

We see that, within the errors, our results are compatible with the 
experimental results given in Eqs.~(\ref{phiexp}) and (\ref{philep}).
Therefore, in the case of the $D_s$ decay into $\phi$, we can say that the 
factorization approximation works well. From this it seems reasonable
to suppose that, if $f_0(980)$ has a dominant $\bar{s}s$ component, the
factorization approximation should also work well for the $D_s$ decay
into $f_0(980)$. Using Eqs.~(\ref{wid}), (\ref{amf0}) and (\ref{amphi}), 
and the values for the form factors at $t=0$ given in Eq.(\ref{fornum}) we 
get:
\beq
{\Ga(D_s^+\rightarrow f_0(980)\pi^+)\over \Ga(D_s^+\rightarrow 
\phi\pi^+)}=0.44\pm0.18\;.
\label{fin}
\enq

In the recently measured spectra from the reaction $D_s^+\to\pi^+\pi^+\pi^-$
\cite{Ait01}, the relative weight of the channel $\pi^+f_0(980)$ is evaluated
to be
\beq
{{\cal B}(D_s^+\to\pi^+f_0(980)){\cal B}(f_0(980)\to\pi^+\pi^-)
\over{\cal B}(D_s^+\to\pi^+\pi^+\pi^-)}=(56.5\pm6.4)\%\;,
\lb{pif0}
\enq
and the ratio of yields
\beq
{\Ga(D_s^+\rightarrow \pi^+\pi^+\pi^-)\over \Ga(D_s^+\rightarrow 
\phi\pi^+)}=0.245\pm0.028^{+0.019}_{-0.012}\;,
\label{raex}
\enq
is measured. Taking into account the results in Eqs.~(\ref{pif0}) and 
(\ref{raex}) one gets:
\beq
R={\Ga(D_s^+\rightarrow f_0(980)\pi^+)\over \Ga(D_s^+\rightarrow 
\phi\pi^+)}={0.140\pm0.046\over{\cal B}(f_0\to\pi^+\pi^-)}\;.
\label{finex}
\enq

Using the branching ratio ${\cal B}(f_0(980)\to
\pi^+\pi^-)\simeq53\%$, the authors in ref.~\cite{adn} have estimated
the ratio $R$ to be $R=0.275(1\pm0.25)$.
In a different way E791, using couple-channel Breit-Wigner function 
\cite{Ait01},
found a non significative $g_K$, that means indirectly a non 
significative contribution for the decay channel  $f_0(980)\to K K $. 
Thus if  we  assume that ${\cal B}(f_0(980)\to\pi\pi)\sim1$ which implies
${\cal B}(f_0(980)\to\pi^+\pi^-)\sim2/3$ (2/3 being the isospin factor),
using this  in Eq.~(\ref{finex}) we get 
\beq
{\Ga(D_s^+\rightarrow f_0(980)\pi^+)\over \Ga(D_s^+\rightarrow 
\phi\pi^+)}=0.210\pm0.069\;.
\label{fin2}
\enq
Therefore, from our result in Eq.~(\ref{fin}), we conclude that there
is a significant deviation from the factorization approximation
for the $D_s^+\rightarrow f_0(980)\pi^+$ decay. This could be an indication 
that there is a sizeable nonstrange component in the $f_0(980)$ meson, or 
even that the $f_0(980)$ structure is more complex than indicated by the 
naive quark model. 

It is interesting to notice that the result obtained
in ref.~\cite{detc} for the ratio in Eq.~(\ref{rat}) is very similar
to our result in Eq.~(\ref{fin}). However, the authors in \cite{detc}
concluded that their result supports a description of $f_0(980)$ as 
a $s\bar{s}$ state with a possible virtual $K\bar{K}$ cloud, but
with no substantial mixture of $u\bar{u},~d\bar{d}$. We believe that
this conclusion was reached because the authors in ref.~\cite{detc}
misinterpreted the experimental result \cite{Ait01}. In their words,
the E791 collaboration measured $R=0.62$,
with a very small error. Since from the Particle Data Group \cite{PDG} we 
only know that
${\cal B}(f_0\to\pi\pi)$ is dominant without knowing the exact number,
there is still an indetermination in the ratio Eq.~(\ref{finex}). As 
explained above,
if we consider ${\cal B}(f_0\to\pi^+\pi^-)\sim2/3$, we arrive at 
the result in Eq.~(\ref{fin2}), which is  smaller than our result in 
Eq.~(\ref{fin}),
leading us to an opposite conclusion compared with \cite{detc}.

One possible way to test if there really is a sizeable nonstrange component 
in the $f_0(980)$ is through the measurement of the semileptonic
$D_s^+\to f_0(980)\ell^+\nu$ decay, since in this decay we do
not have problems with the factorization approximation. Our predicition
for the branching ratio obtained from Eq.(\ref{lepf0}),
by supposing $f_0(980)$ as a $\bar{s}s$ state, is: 
\beq
{\cal B}(D_s^+\to f_0(980)\ell^+\nu)= (0.55 \pm 0.10)\% \;.
\enq
Any significant deviation from that will definitively imply in a
sizeable nonstrange component in the $f_0(980)$ meson, which could be
or not accomodated in the naive quark model. Therefore, we urge the 
experimentalists to search for this decay.

\section{Summary and conclusions}

We have presented a QCD sum rule study of the $D_s^+$ decays to final
states containing $\phi$ and $f_0(980)$ mesons. We have evaluated the $t$
dependence of the form factors
$f_+(t)$, $A_1(t)$, $A_2(t)$ and $V(t)$ in the region $-0.5\leq t\leq0.4
\GeV^2$.
The $t$ dependence of the form factors could be fitted by a monopole form
and extrapolated to the full kinematical region. The axial-vector form
factors $A_1$ and $A_2$ have a much weaker $t$ dependence than the form
factors $f_+$ and $V$.

The form factors were used to evaluate the branching ratios for
the decays $D_s^+\to\phi\pi^+$ and $D_s^+\to\phi\ell^+\nu_\ell$
and we have obtained a good agreement with experimental data. 
Since the evaluation of the decay
width, in the nonleptonic decay, is based on the factorization approximation,
our first conclusion is that the factorization approximation works well
in the case of the decay $D_s^+\to\phi\pi^+$. 

We have also evaluated the ratio $D_s^+\to f_0(980)\pi^+/(D_s^+\to\phi\pi^+)$
and we got a result bigger than estimate based on experimental data.
Based on the fact that factorization approximation works well
in the case of the decay $D_s^+\to\phi\pi^+$, this result can be interpretated
as an indication that
there is a sizeable nonstrange component in the $f_0(980)$ meson, or even
that the $f_0(980)$ structure is more complex than indicated by the naive
quark model. This hypothesis could be tested by the measurement of the
semileptonic $D_s^+\to f_0(980)\ell^+\nu$ decay, since there is no problem
with the factorization approximation in the semileptonic decays.

\vspace{1cm}
 
\underline{Acknowledgements}: 
We would like to thank H.G. Dosch for giving us the idea for this 
calculation and F.S. Navarra for fruitful discussions.  
This work has been supported by CNPq and  FAPESP (Brazil). 
\vspace{0.5cm}

\appendix

\section{Two-point sum rules}

In ref.~\cite{faz} the two-point sum rule for the $f_0(980)$ meson
was evaluated by considering $f_0$ as a $\bar{s}s$ state. They got:
\beqa
m_{f_0}^2f_{f_0}^2e^{-m_{f_0}^2/M^2}&=&{3\over8\pi^2}\int_{4m_s^2}^{u_0}
du u\left(1-{4m_s^2\over u}\right)^{3/2}e^{-u/M^2}\nn\\
&+&m_se^{-m_s^2/M^2}\left[\lag\bar{s}s\rag\left(3+{m_s^2\over M^2}\right)
+{\lag\bar{s}g_s\si.Gs\rag\over M^2}\left(1-{m_s^2\over2M^2}\right)\right]
\;.
\enqa

Considering $M^2$ in the interval $1\leq M^2\leq2\GeV^2$, $u_0=1.6\pm0.1\GeV^2$
 they got 
\beq
f_{f_0}=(0.180\pm0.015)\GeV.
\lb{f0nu}
\enq
If we consider $m_s^2=0$, the result for $f_{f_0}$ does not change 
significantly, and we get $f_{f_0}=(0.19\pm0.02)\GeV$. 

The sum rule for $\phi$ is given by \cite{rr}:
\beq
f_\phi^2e^{-m_\phi^2/M^2}={1\over4\pi^2}\int_{4m_s^2}^{u_0}
du {(u+2m_s^2)\sqrt{u-4m_s^2}\over u^{3/2}}
e^{-u/M^2}+{2m_s\lag\bar{s}s\rag\over M^2}+{\lag g^2G^2\rag\over
48\pi^2M^2}.
\enq
Considering $m_s$ at most linearly and using $u_0=2.0\pm0.1\GeV^2$ we get
\beq
f_{\phi}=(0.232\pm0.010)\GeV,
\lb{phinu}
\enq
in the interval $1\leq M^2\leq2\GeV^2$, in a very good agreement with the 
experimental value $f_\phi^{exp}=0.234\GeV$ \cite{PDG}.

For $f_{D_s}$ the two point sum rule is given by:
\beqa
&&{m_{D_s}^4f_{D_s}^2\over(m_c+m_s)^2}e^{-m_{D_s}^2/M^2}={3\over8\pi^2}
\int_{(m_c+m_s)^2}^{s_0} ds
(1-{(m_c-m_s)^2\over s}\bigg)\sqrt{\la(s,m_c^2,m_s^2)}e^{-s/M^2}\nn\\ 
&+&\lag\bar{s}s\rag e^{-m_c^2/M^2}\left(-m_c+{m_s\over2}+{m_sm_c^2\over M^2}
\right)-{m_c\lag\bar{s}g_s\si.Gs\rag\over 2M^2}e^{-m_c^2/M^2}\left(1-
{m_c^2\over2M^2}\right)
\;.
\enqa
Considering $m_s$ at most linearly and using $s_0=8.8\GeV^2$ we get
\beq
f_{D_s}=(0.22\pm0.02)\GeV,
\lb{dsnu}
\enq
in the interval $2.3\leq M^2\leq4.5\GeV^2$, in a good agreement with the 
value quoted in ref.~\cite{ck00}, and also with recent lattice
determination \cite{lat}: $f_{D_s}=(0.252\pm0.009)\GeV$.

\section{Perturbative Contributions to the three-point functions}

In all this work we take into account the mass of the strange quark at most 
linearly. We have checked that the contibution of terms proportional to 
$m_s^2$
and higher powers are negligible. The perturbative 
contributions for the sum rules defined in Sec.~III are:
\beqa
\rho_+(s,u,t)&=&{3\over\la^{3/2}(s,u,t)}\left\{u\left[2m_cm_s(2m_c^2-s-t+u)
+m_c^2(s-t+u)+s(-s+t+u)\right]\right.\nonumber\\
&-&\left.(2m_c^2-s-t+u)(su+m_cm_s(s-t+u))\right\}\theta(s-s_M),
\enqa
\beqa
\rho_1(s,u,t)&=&-{3\over2\la^{3/2}(s,u,t)}\left\{m_cu\left[\la(s,u,t)+
2m_c^2(m_c^2-s-t+u)+2st\right]\right.
\nonumber\\
&+&m_s\left[s^3-t^3+2ut^2+ut(2m_c^2-u)-2m_c^2u(m_c^2+u)\right.
\nonumber\\
&-&\left.\left.s^2(3t+2u)+s(3t^2-
2tu+u(2m_c^2+u))\right]\right\}\theta(s-s_M),
\enqa
\beqa
\rho_2(s,u,t)&=&{3\over2\la^{5/2}(s,u,t)}\left\{m_cu(s-t-u)\left[\la(s,u,t)
+6st+6m_c^2(m_c^2-s-t+u)\right]\right.
\nonumber\\
&+&m_s\left[s^4
+t^4-t^3u+2m_c^2u^2(3m_c^2+u)-ut^2(10m_c^2+u)\right.
\nonumber\\
&-&s^3(4t+3u)+tu(6m_c^4+8m_c^2u+
u^2)+s^2(6t^2-tu+u(2m_c^2+3u))
\nonumber\\
&-&\left.\left.s(4t^3-5t^2u-4tu(2m_c^2+u)+u(6m_c^4+4m_c^2u+
u^2))\right]\right\}\theta(s-s_M),
\enqa
\beqa
\rho_V(s,u,t)&=&{3\over2\la^{3/2}(s,u,t)}\left\{m_cu\left(
2m_c^2-s-t+u\right)\right.
\nonumber\\
&+&\left.
m_s\left[s^2+t^2-ut-2um_c^2-s(2t+u)\right]\right\}\theta(s-s_M),
\enqa
where 
\beq
s_M=m_c^2+{m_c^2u\over m_c^2-t}.
\enq


\end{document}